\input{aipcheck}

\documentclass[
    ,final            
  ]
  {aipproc}

\def\etal{et al.~}
\def\Nature{{\it Nature}}
\def\ApJ{{\it Astrophys. J.}}
\def\MNRAS{{\it Mon. Not. R. Astron. Soc.}}
\def\AA{{\it Astron. Astrophys.}}

\layoutstyle{8x11single}

\begin{document}

\title{Magnetized GRB outflow model:  weak reverse shock emission and short energy transfer
timescale\footnote{My presentation is based on arXiv:0805.2221, a
review article published in {\it Front. Phys. China.} The current
paper focuses on the 17th slide of that PPT
(http://grb.physics.unlv.edu/nj/talks/6.26/$\rm
Fan_{-}Yizhong.ppt$).}}

\author{Yi-Zhong Fan$^{1,2}$} {
  address={$^1$ Niels Bohr International Academy, Niels Bohr
  Institute, Copenhagen University, Blegdamsvej 17, DK-2100 Copenhagen,
          Denmark.\\
          $^2$ Purple Mountain Observatory, Chinese Academy of
Sciences, Nanjing 210008, China. }}

\begin{abstract}
We show that the absence of the bright optical flashes in most {\it Swift} Gamma-Ray
Burst (GRB) afterglows can be explained, if the reverse shock region is magnetized with a
$\sigma \sim 1$, or the emission spectrum of the electrons accelerated in the mildly
magnetized ($0.1<\sigma<1$) reverse shock front is very soft, or the reverse shock of a
non-magnetized outflow is sub-relativistic, where $\sigma$ is the ratio of the magnetic
energy flux to the particle energy flux. We also find that for $\sigma\gg 1$, the energy
transfer between the magnetized ejecta and the forward shock may be too quick to account
for the shallow decline phase that is well detected in many {\it Swift} GRB X-ray
afterglows.
\end{abstract}

\maketitle


\noindent%
\section{Magnetized GRB outflow model}
Though extensively discussed, the physical composition of the GRB
outflow is not clear yet (see \cite{Piran04,Zhang07} for reviews).
In principle, the outflows could be either Poynting-flux dominated
or baryon dominated. The former is favored if the central engine is
a millisecond magnetar \cite{Usov94} or the outflow is launched from
a black hole$-$torus system through MHD processes. For a magnetized
outflow, the prompt $\gamma-$rays are powered by the magnetic energy
dissipation \cite{Usov94,LB03,GS06} or magnetized internal shocks
\cite{FWZ04}. A signature is the high linear polarization of the
prompt emission \cite{LPB03,G03}, which was reported \cite{CB03}
 in GRB 021206 but afterwards ruled out \cite{RF04}. In a
few other events, high linear polarization has been claimed  but
independent measurements for each burst are needed to confirm these
discoveries.

A ``robust" evidence for the magnetized GRB outflow model may be the magnetization of the
GRB reverse shock (RS). Shortly after the discovery of the very bright optical flash of
GRB 990123 \cite{Akerlof99}, Sari \& Piran \cite{SP99} and M\'esz\'aros \& Rees
\cite{MR99} showed that the $t^{-2}$ decline of the optical flash can be well interpreted
by the adiabatically cooling of the RS electrons. The self-consistent fitting of the very
early and the late time afterglow data requires that $\epsilon_B$, the fraction of the
shock energy given to the magnetic field, of the RS is much larger than that of the
forward shock (FS) \cite{Fan02,ZKM03,PK04}. It is very interesting to note that such a
finding has been confirmed in almost all optical flash modelings, such as for GRB 021211
\cite{Fox03,Li03,ZKM03,KP03,PK04}, GRB 041219a \cite{Blake05,FZW05}, GRB 050401
\cite{Blustin06}, GRB 050904 \cite{Boer06,Wei06}, GRB 060111B \cite{Klotz06}, GRB 061126
\cite{Gomboc08}, and GRB 080319B \cite{Racusin08}.
A natural interpretation for this finding is that the GRB outflow is magnetized and the
magnetic field in the RS region is dominated by the component carried from the central
engine.

The magnetized outflow model also helps to solve the puzzle why
optical flashes have not been detected in most {\it Swift} GRBs
\cite{Roming06}. This is because for a highly magnetized GRB
outflow, under the ideal MHD limit, the magnetic energy can not be
converted into the RS energy effectively. The emission of a
magnetized GRB RS has been calculated by Fan et al. \cite{FWW04} and
Zhang \& Kobayashi \cite{ZK05} (see also \cite{G07} for the
existence of such shocks). With reasonable parameters ($\sigma<10$),
the resulting RS emission is very bright and is inconsistent with
the observational data. It is the time to revisit these preliminary
calculations.

\section{Weak reverse shock emission}
Following \cite{FWW04}, we calculate the emission of magnetized RSs.
Some important modifications include: (i) The arrival time of the RS
emission photons has been calculated more accurately (see the
Appendix for details). (ii) The electron energy distribution is
calculated by solving the continuity equation with the power-law
source function $Q=K\gamma_e^{-p}$, normalized by a local injection
rate \cite{FP06}. (iii) The cooling of the electrons, due to both
synchrotron and inverse Compton radiation, has been
considered\footnote{Sometimes the overlapping of prompt $\gamma-$ray
flow with the RS region is very tight. In such cases the RS
electrons will be cooled by the prompt $\gamma-$rays, too. As a
result, the RS optical emission will be dimmer than what we get in
this work.}. (iv) The synchrotron self-absorption has also been
taken into account.

\begin{figure}[h]
\includegraphics*[width=.7\textwidth]{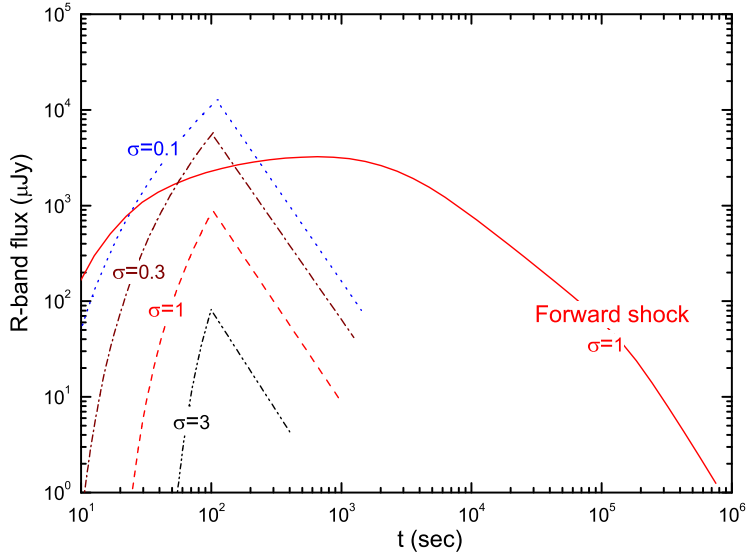}
\caption{The reverse shock optical emission of the outflows with
magnetization. Except the $\sigma$ marked in the figure, other
physical parameters are taken as: the duration of the burst
$T_{90}=100$ s, the initial Lorentz factor $\Gamma_0=300$, $z=1$,
$E_{\rm k}=10^{53}/(1+\sigma)$ erg, $n=1~{\rm cm^{-3}}$, $p=2.3$,
and $\epsilon_e=0.1$. For the FS, we assume $\epsilon_B=0.01$.}
\label{fig:sigma_wide}
\end{figure}
We find a bright (RS) optical flash is absent in the following scenarios:

\textbf{\emph{Significant magnetization.}} After the important
corrections mentioned above, we find a $\sigma \sim 1$ is enough to
suppress the RS emission effectively and thus be able to explain the
absence of the bright optical flashes in most GRB afterglows (see
Fig.\ref{fig:sigma_wide}). This renders the magnetization
interpretation more attractive because $\sigma$ declines with radius
after the prompt emission and is expected to be $<10$ at a radius $>
10^{16}$ cm \cite{Usov94,GS06}.

For clarity, in Fig.\ref{fig:sigma_wide} we plot the FS emission for
$\sigma=1$ while the RS emission is presented for
$\sigma=(0.1,~0.3,~1,~3)$, respectively. The $\epsilon_e$ values are
assumed to be the same for both RS and FS. The new results are
significantly different from those obtained in \cite{FWW04,ZK05}.

\textbf{\emph{Mild magnetization but a very soft spectrum of the RS
emission.}} In all previous calculations, the energy distribution
index, $p$, of the RS electrons is taken to be the same as that of
the FS electrons. But this treatment may be well wrong. We know that
usually $p$ is relevant to $(\beta_{\rm u}-\beta_{\rm d})^{-1}$,
where $\beta_{\rm u}$ and $\beta_{\rm d}$ are the velocities (in
units of the speed of light $c$ and measured in the rest frame of
the shock front) for the upstream and the downstream regions,
respectively. For a relativistic un-magnetized shock, we have
$\beta_{\rm u} \sim 1$, $\beta_{\rm d} \sim 1/3$, and $(\beta_{\rm
u}-\beta_{\rm d})^{-1}\approx 3/2$. However, for a relativistic
magnetized shock, $\beta_{\rm d} \approx {1\over
6}(1+\chi+\sqrt{1+14\chi+\chi^2})$, where $\chi\equiv
\sigma/(1+\sigma)$ \cite{FWZ04}. For $\sigma\sim {\rm a ~few}\times
0.1$, $(\beta_{\rm u}-\beta_{\rm d})^{-1} \approx 3/(2-4\sigma)$,
{\it the energy distribution of the RS electrons may be much steeper
than in the case of $\sigma=0$} \footnote{A speculation is that some
prompt $\gamma-$ray emission with a very soft spectrum might be
powered by the magnetized internal shocks and should be linearly
polarized.}. There are two indication evidences for this
speculation: (i) In a numerical calculation, Morlino et al.
\cite{Morl07} found a $p\sim 3$ for $\sigma \sim 0.05$, and (ii) In
the 2.5D ion-electron shock simulation, the acceleration of
particles in the case of $\sigma \sim 0$ is much more efficient than
in the case of $\sigma \sim 0.1$ \cite{Spit06}.

Assuming that the RS electrons have a $p\sim p_{(\sigma=0)}+\Delta
p$, the RS optical emission will be weakened by a factor of ${\cal
R}_{\rm w}\sim (\nu_{\rm opt}/\nu_{\rm m}^{\rm rs})^{-{\Delta p
\over 2}}$ as long as $\nu_{\rm opt}>\nu_{\rm m}^{\rm rs}$, where
$\nu_{\rm opt}\sim 4\times 10^{14}$ Hz is the observer's frequency
and $\nu_{\rm m}^{\rm rs}$ is the typical synchrotron radiation
frequency of the RS electrons. For $\nu_{\rm m}^{\rm rs}\leq
0.1\nu_{\rm opt}$ and $\Delta p \sim 1$, we have ${\cal R}_{\rm w}
\leq 0.3$. Such a correction will render the RS optical emission for
$\sigma=0.3$, as shown in Fig.\ref{fig:sigma_wide}, outshone by the
FS emission. As a result, a bright RS optical flash is absent.

\begin{figure}[h]
\includegraphics*[width=.7\textwidth]{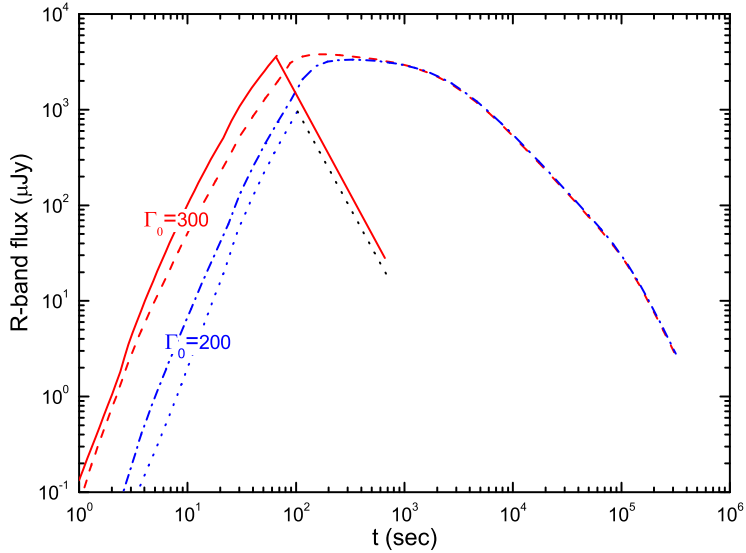}
\caption{The RS emission in the case of $\sigma=0$. The solid and
the dotted lines are the RS emission component. The dashed and the
dash-dotted lines are the FS emission. All parameters, except
$T_{90}=60$ s and $E_{\rm k}=10^{53}$ erg, are the same as those in
Fig.\ref{fig:sigma_wide}.} \label{fig:sigma_zero}
\end{figure}

\textbf{\emph{Non-magnetization but a very weak RS.}} Nakar \& Piran
\cite{NP04} and Jin \& Fan \cite{JF07} also got very weak RS optical
emission in the case of $\sigma=0$. Here we investigate the
influence of the strength of the RS on the peak optical emission in
such a particular case. The numerical results have been presented in
Fig.\ref{fig:sigma_zero}. As expected, the stronger the RS, the
brighter the optical emission. The very weak RS emission implied by
the {\it Swift} UVOT observation strongly suggests a
sub-relativistic RS provided that $\sigma=0$.

\section{Magnetic energy transfer timescale} In the very early afterglow phase
(during which both the RS and FS exist), the magnetic energy of the
outflow can not be converted into the kinetic energy of the FS
effectively. After the ceasing of the RS, a significant energy
transfer between the magnetized outflow and the FS is possible. In
about half of {\it Swift} GRB X-ray afterglows, a long term
flattening is evident. This phenomenon motivates an idea that the
magnetic energy has been transferred into the kinetic energy of the
FS continually but slowly and then gives rise to a shallow X-ray
decline phase \cite{Zhang06}. If correct, the magnetized outflow
model will be strongly favored because it can also naturally account
for the absence of the bright optical flashes in most {\it Swift}
GRBs (see the previous section). It is thus highly needed to
calculate the transfer timescale of the magnetic energy.

\begin{figure}[h]
\includegraphics*[width=.7\textwidth]{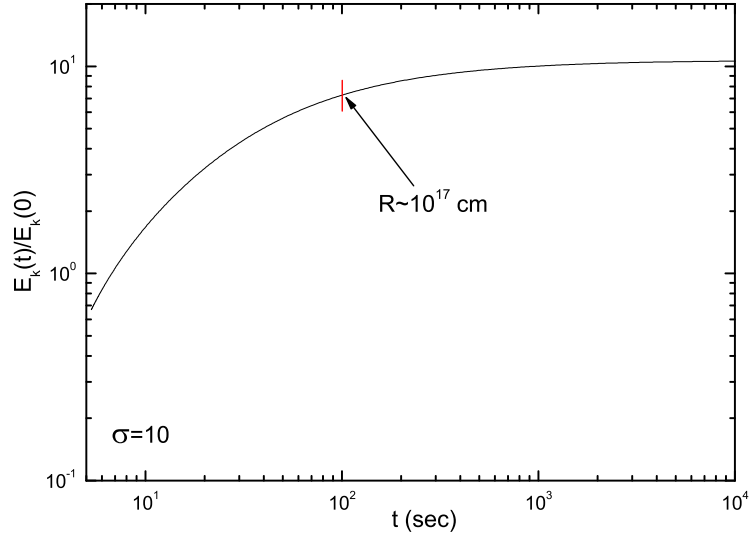}
\caption{The kinetic energy of the FS driven by a magnetized outflow
($\sigma=10$) as a function of the observer's time. One can see that
at an observer's time $\sim 100$ s, most magnetic energy has been
converted into the kinetic energy of the FS.} \label{fig:transfer1}
\end{figure}

A preliminary investigation has been carried out by Fan \& Piran in
\cite{FP06} (see \S{3.5} therein). The basic idea is that: For a
highly magnetized outflow, the magnetic energy can not be converted
to the kinetic energy of the ejecta in a single passage of the
reverse shock. There is a possibility to form multiple RSs as long
as the total pressure behind the contact discontinuity is lower than
the thermal pressure of the shocked medium. If the total pressure
behind the contact discontinuity gets higher, the RS ceases and the
magnetic pressure works upon the shocked medium and leads to the
increase of the FS's energy. The deceleration of the FS is thus
suppressed and the afterglow light curves are flattened. The energy
transfer timescale, however, seems to be too short (see
Fig.\ref{fig:transfer1} for illustration) to account for the shallow
decline phase lasting $\sim 10^4$ sec that is detected in many {\it
Swift} X-ray afterglows. The very recent numerical simulations
confirm our conclusions \cite{Mizu08,Mimi08}.

\section{Conclusions}
In this work, we find that:
\begin{itemize}
\item The magnetization of the GRB outflows plays a crucial role in
suppressing the reverse shock optical emission. Bright optical
flashes are only expected in the case of $\sigma<0.1$. Polarimetry
of the optical flashes should have significant detections.

\item The energy transfer between a highly magnetized ejecta and the
forward shock may be too quick to account for the shallow decline phase that is detected
in a good fraction of {\it Swift} GRB X-ray afterglows.
\end{itemize}

\begin{theacknowledgments}
I thank Tsvi Piran, Da-Ming Wei and Bing Zhang for fruitful
collaborations. This work was supported by Danish National Science
Foundation and Chinese Academy of Sciences.
\end{theacknowledgments}

\begin{appendix}
\section{The propagation of photons: shaping the
reverse shock emission light curve}
\begin{figure}[h]
\includegraphics*[width=.7\textwidth]{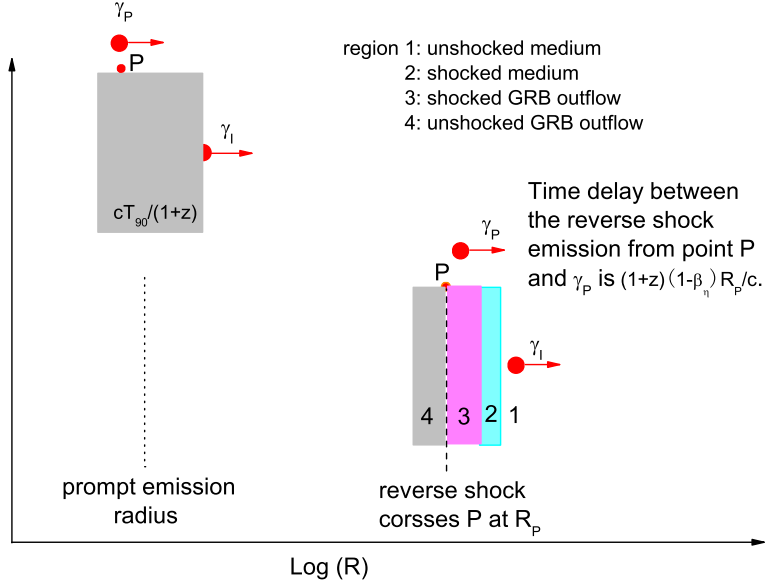}
\caption{The arrival time of the RS emission.} \label{fig:illust1}
\end{figure}
At a radius $R_\times$, the RS crosses the GRB ejecta with a width
$\Delta \simeq cT_{90}/(1+z)$.  The RS ``crossing time" is estimated
as
\begin{equation}
t_\times=(1+z)\int^{R_\times}_0 (1-\beta_{\rm \Gamma_3}){dR\over c},
\end{equation}
where $dR=cdt/(\beta_{\eta}-\beta_{\rm rsh})$, $\Gamma_3$ and $\eta
(\equiv \Gamma_0)$ are the bulk Lorentz factors of regions 3 and 4,
respectively (measured in the observer's frame), and $\beta$ is the
corresponding velocity in units of $c$. With the relation
$\Gamma_{\rm rsh}\approx (\gamma_3-u_3)\Gamma_3$, we have
\cite{FWW04}
\begin{equation}
t_\times \approx \frac{1}{(\gamma_{3,\times}+u_{3,\times})^2}T_{90},
\end{equation}
where $\gamma_3$ is the Lorentz factor of the shocked fluid
(measured in the reverse shock frame) and $u_3=\sqrt{\gamma_3^2-1}$.
In the case of $\sigma=0$ and the RS is relativistic, we have
$\gamma_{3,\times}\approx \sqrt{9/8}$ and $t_\times \approx T_{\rm
90}/2$. While for $\sigma \gg 1$, we have $\gamma_{3,\times}\approx
\sqrt{\sigma}$ and $t_\times \approx T_{\rm 90}/4\sigma$
\cite{FWW04}\footnote{The magnetized RS does have a shorter crossing
time, as confirmed by the numerical simulation \cite{Mimica08}.}.
{\it So the simple treatment that takes $t_\times$ as the RS
emission timescale may violate the causality and thus be flawed.}

We then have to look for a more reliable calculation of the arrival
time of the RS emission (see also \cite{Wei06}). The zero point of
the observer's time is that of the first $\gamma$-ray photon we
detected. For illustration, we consider the case of $\theta=0$
(i.e., on the line of sight), a $\gamma$-ray photon ${\gamma}_{\rm
P}$ arriving at $t_{\rm em,P}$ implies that the distance from the
corresponding electron (i.e., point P, at which the bulk lorentz
factor is $\eta$) to the initial outflow front is $\approx ct_{\rm
em,P}/(1+z)$. The radial distance from the FS front to the central
engine is $R_{\rm P}$ when the RS crosses point $P$. At that time,
the separation between photon
 ${\gamma}_{\rm P}$ and point $P$ is $\approx (1-\beta_\eta)R_{\rm P}$.
 Therefore, the arrival
 time of the RS emission from point $P$ should be (see
 Fig.\ref{fig:illust1} for illustration)
 \begin{equation}
 t_{\rm arr,P}\sim
 t_{\rm em,P}+(1+z)(1-\beta_\eta)R_{\rm P}/c.
 \end{equation}
If $P$ is the rear of the GRB outflow, $t_{\rm arr,P}$ should be
always larger than $T_{90}$. So, {\it in Fig.1 and Fig.2 of
\cite{FWW04}, the RS emission duration has been underestimated and
the flux overestimated significantly because the total energy
emitted in that phase is fixed.}

The crossing radius for point $P$ can be calculated as $R_{\rm P}
\approx 2 \Gamma_{\rm rsh,P}^2 c t_{\rm em,P}/(1+z)$. Similarly, for
point $Q$, $R_{\rm Q} \approx 2 \Gamma_{\rm rsh,Q}^2 c t_{\rm
em,Q}/(1+z)$. The corresponding crossing times are $t_{\rm
\times,P}\approx t_{\rm em,P}\Gamma_{\rm rsh,P}^2/\Gamma_{3,P}^2$
and $t_{\rm \times,Q}\approx t_{\rm em,Q}\Gamma_{\rm
rsh,Q}^2/\Gamma_{3,Q}^2 $, respectively.
The arrival times, however, are $t_{\rm arr,P}
\sim t_{\rm em,P}$ and $t_{\rm arr,Q} \sim t_{\rm em,Q}$,
respectively since both $\Gamma_{\rm rsh,P}^2/\eta^2$ and
$\Gamma_{\rm rsh,Q}^2/\eta^2$ are $\ll 1$. Following \cite{Koba00},
we can calculate the RS emission. However, the resulting light
curves are scaled by the ``crossing" time, which should be
transferred into what are scaled by the arrival time. Because of the
energy conservation, for $P \rightarrow Q$, the emissions are
related by ${F_{\nu_{\rm opt}(t_{\rm \times,P})}+F_{\nu_{\rm
opt}(t_{\rm \times,Q})} \over 2}(t_{\rm \times,Q}-t_{\rm \times,P})
\approx {F_{\nu_{\rm opt}(t_{\rm arr,P})}+F_{\nu_{\rm opt}(t_{\rm
arr,Q})} \over 2}(t_{\rm arr,Q}-t_{\rm arr,P})$, which yields
\begin{equation}
F_{\nu_{\rm opt}(t_{\rm arr,P})} \sim {t_{\rm \times,P} \over t_{\rm
arr,P}} F_{\nu_{\rm opt}(t_{\rm \times,P})}.
\end{equation}
\end{appendix}

\end{document}